\documentclass[9pt]{extarticle}
\usepackage{spconf,amsmath,graphicx}
\usepackage{graphicx}
\usepackage{amsmath}
\usepackage{amssymb}
\usepackage{booktabs}
\usepackage{caption}
\usepackage{subcaption}
\usepackage[pagebackref,breaklinks,colorlinks]{hyperref}

\title{Xi-Net: Transformer based Seismic Waveform Reconstructor}
\name{Anshuman Gaharwar\textsuperscript{1,}*\thanks{* Equal contribution}, Parth Parag Kulkarni\textsuperscript{1,}*, Joshua Dickey\textsuperscript{2}, Mubarak Shah\textsuperscript{1}}
\address{\textsuperscript{1}Center for Research in Computer Vision, University of Central Florida\\\textsuperscript{2}Air Force Technical Applications Center(AFTAC), US Air Force}

\begin{document}
%
\maketitle
\begin{abstract}Missing/erroneous data is a major problem in today’s world. Collected seismic data sometimes contain gaps due to multitude of reasons like interference and sensor malfunction. Gaps in seismic waveforms hamper further signal processing to gain valuable information. Plethora of techniques are used for data reconstruction in other domains like image, video, audio, but translation of those methods to address seismic waveforms demands adapting them to lengthy sequence inputs, which is practically complex. Even if that is accomplished, high computational costs and inefficiency would still persist in these predominantly convolution-based reconstruction models. In this paper, we present a transformer-based deep learning model, Xi-Net, which utilizes multi-faceted time and frequency domain inputs for accurate waveform reconstruction. Xi-Net converts the input waveform to frequency domain, employs separate encoders for time and frequency domains, and one decoder for getting reconstructed output waveform from the fused features. 1D shifted-window transformer blocks form the elementary units of all parts of the model. To the best of our knowledge, this is the first transformer-based deep learning model for seismic waveform reconstruction. We demonstrate this model's prowess by filling 0.5-1s random gaps in 120s waveforms, resembling the original waveform quite closely. The code, models can be found \href{https://github.com/Anshuman04/waveformReconstructor}{here}.
\end{abstract}
\begin{keywords}
Xi-Net, Seismic, Transformer, Reconstruction, Frequency 
\end{keywords}
\section{Introduction}
\label{section:intro}

Signals collected from the geophysical sensors, are often corrupted or a part of the signal is lost during its transmission to the target site. Such a phenomenon with large gaps equivalent to a second in a 120-second waveform is observed in seismic data collected from various places across the world. Our goal in this work is to efficiently reconstruct the missing part of a waveform using deep-learning based methods.
Solving the problem is not trivial as the missing gaps are large and existing methods of Image Inpainting 
\cite{zhao2021large,zhou2021transfill,yu2021diverse,pirnay2021inpainting,saharia2021palette}
are not feasible as the typical image based representation of the waveform (2D plot as an image input) contains less than 1\% missing pixels. Moreover, existing 1D signal based reconstructions approaches act on small gaps and do not scale.
To the best of our knowledge, our work is the first towards using transformers in vision for this problem. Transformers are better suited for this task over CNNs, owing to the self-attention mechanism for long term dependencies and positional embeddings. Our approach is inspired from Swin-Unet \cite{cao2021swin}, which employs encoder-decoder style based Swin transformer blocks in 2D Medical Image Segmentation. 
In this paper, we present our approach named Xi-Net, a novel pure-transformer based network consisting of encoder-decoder style based Swin transformer blocks to reconstruct gaps in seismic waveforms. Our network relies on waveforms represented in time and frequency domains with separate encoders for each and a single decoder fusing  with skip connections in U-Net style from both encoders to reconstruct the signal. 
Decoder blocks are based on Swin transformer blocks with Patch partition and 1D Patch Merging. 
Given an input waveform, we perform noise removal and upsampling to a dimension of 14400 resulting in timeDomain signal. 
Discrete Time Fourier Transform is applied to the up-sampled time signal resulting in real and imaginary planes of this signal. These planes are laterally stacked to form frequency domain input. Both time and frequency domain inputs are then fed to Xi-Net for processing. 

Our main contributions  are  summarized as follows:
\begin{itemize}
    \item A novel transformer architecture for reconstructing seismic waveforms with missing gaps.
    \item Introduction of 1D window shifting +1D patch merging.
    \item Multi domain reconstruction by fusing time and frequency domain patterns.
    \item Attention mechanism for complex numbers.
\end{itemize}

\begin{figure*}
  \centering
  \includegraphics[width = \textwidth]{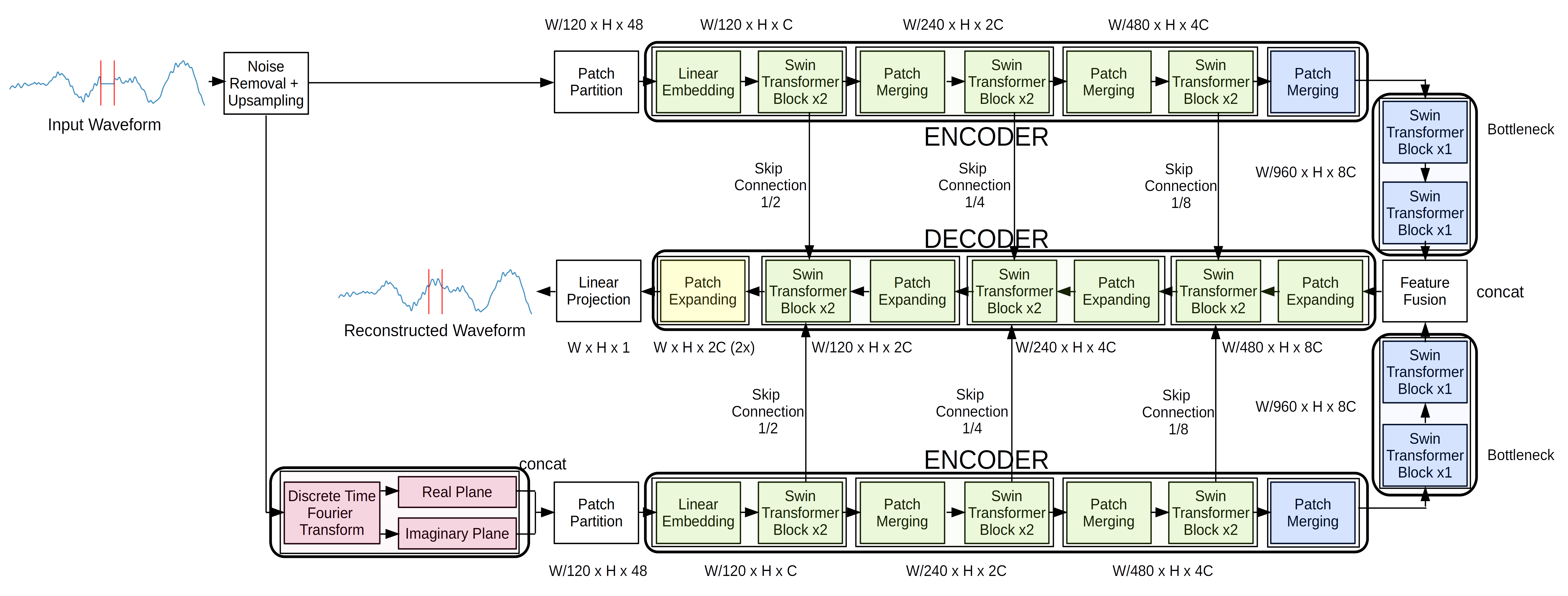}
  \caption{\textbf{Schematic Illustration of the proposed Xi-Net Model Architecture.} The upper branch of the model is the encoder which processes on time domain data. The lower branch is an encoder processing on frequency domain data. The middle branch is the decoder which combinedly processes the multi-channel data to output a reconstructed waveform. An Input waveform goes through upsampling and noise removal before going to the time encoder. The processed waveform also goes through an additional DTFT layer, divider and lateral stacking before going to the frequency encoder. [Here, W and H are the width and the height of the signal respectively. In our case, H=1(time series)]}
  \label{fig:xinet}
\end{figure*}

\section{Related Work}
\label{section:related}
Encoder-decoder models have been the basis behind many networks in various areas of Computer Vision. Xia et. al. \cite{xia2017w} 
proposed a double U-net architecture for image segmentation. Lan et. al. \cite{lan2019net} utilize two encoders and one decoder for image reconstruction. Wang et. al. \cite{wang2019net} also use a similar idea for holographic reconstruction. Cao et. al.\cite{cao2021swin} employ this idea for reconstruction, by using Swin blocks as components of a  model architecture. Audio and time series reconstruction has been an active area of research for some time. Various deep learning based 1D reconstruction techniques have been employed for generating missing data, most of which like Chang et. al.\cite{chang2019deep} and Hadjeres et. al.\cite{hadjeres2021piano} were CNN based. Transformers have recently entered this domain as well due to their prowess and effectiveness in Vision tasks. Wen et. al.\cite{wen2022transformers} present a comprehensive list of such techniques which use transformers for time series processing  including reconstruction. Lin et. al.\cite{lin2020audiovisual} propose an Audiovisual transformer to facilitate event localization, which uses an  interesting technique for audio processing. Bazin et. al.\cite{bazin2021spectrogram} adapt VQ-VAE-2 architecture to convert spectrograms to codemaps which are inpainted using token-masked Transformers. Works like Défossez\cite{defossez2021hybrid} and Rouard et. al.\cite{rouard2022hybrid} use hybrid temporal-spectral transformer networks, albeit for source separation, instead of reconstruction.

In this paper, we propose a novel solution to the problem of seismic waveform reconstruction. The simultaneous processing of time and frequency versions of the waveforms allows us to achieve good quality reconstructions. 

\section{Xi-Net}
\label{section:xi}
In this section we provide  the detailed architecture of Xi-Net, 
which makes use of both time and frequency domain patterns to reconstruct waveforms.

\subsection{Preprocessing}
\label{section:preprocessing}
In the field of waveform reconstruction, it is well accepted that better results can be achieved by upsampling the input waveforms, as this provides granularity to both pattern detection in waveforms (encoding) and final reconstruction from patterns (decoding). Therefore, we  perform upsampling of the waveform by a factor of 2, as a means of data augmentation. Noise removal  is part and parcel of the signal processing field and there are many different types of noise removal strategies for keeping the reconstruction loss to minimum. We  use Butterworth Bandpass filtering\cite{butterworth1930theory} for noise removal in our dataset. This mechanism of noise removal provides flat frequency response in passband frequencies. Hence, we preprocess the complete dataset by upsampling and noise removal.

\subsection{Frequency Tokenization}
\label{section:freq}
Our proposed architecture consists of two encoders, one each for time and frequency domain analysis. Later, the encoded patterns are used to reconstruct the waveforms by the  decoder. As the decoder  uses patterns from both domains, 
 to ensure that no bias is formed towards either time or frequency domain encoder, we  use the same structure and blocks of Swin for encoding. We 
employ Discrete Time Fourier Transform (DTFT) to convert  time series data $(W \times H)_{Real}$ into frequency domain representation $(W \times H)_{Complex}$. This results in generation of complex numbers which can not be passed to the frequency encoder directly. We propose an alternative representation of complex numbers specifically targeting the self-attention mechanism of transformers. We  separate both real and imaginary values for each complex number resulting in two planes of the same resolution $(W \times H)_{Real}$. Next,  both  planes are stacked on the depth axis to get a new representation $(W \times H \times 2)_{Real}$. This new representation is analogous to having a multi-channel input to transformers (e.g. RGB image when passed to transformers has 3 channels stacked, similarly complex numbers can be represented as 2 channels stacked together, each having real values). This unique representation enables our architecture to capitalize on both time and frequency domain patterns.

\subsection{Model Architecture}
\label{section:arch}
The name Xi-Net is inspired from uppercase Greek letter Xi ($\Xi$). As the symbol has three horizontal lines our architecture has 3 pillars: Time series (T) encoder, Frequency (F) encoder, T+F decoder which is shown in Fig.\ref{fig:xinet} Each logical block from our architecture is explained below:

\subsubsection{Embedding waveforms}
\label{section:ew}
We consider the waveform $W_T \in R^{W \times H \times 1}$ as  input to the architecture. $W_T$ is a time domain waveform which is already pre-processed, i.e $W_T$ is an upsampled waveform and already passed through a Butterworth Bandpass filter. We compute DTFT (as discussed in Section \ref{section:freq}) resulting in its frequency domain representation, $W_F \in C^{W \times H \times 1}$. $W_F$ is further separated into its real ($W_F^R \in R^{W \times H \times 1}$) and imaginary ($W_F^I \in R^{W \times H \times 1}$) counterparts. Both real and imaginary parts are then stacked together in depth producing transformer friendly representation for complex entities, $W_{FF} \in R^{W \times H \times 2}$.  Next, $W_T$ and $W_{FF}$ are passed to their respective encoders through separate patch partition same as \cite{dosovitskiy2020image}.

\subsubsection{Encoders}
\label{section:enc}
The architecture consists of time and frequency encoders, both having the exact same architecture design. The only difference between the two encoders is the dimensions of input  tokens. The input frequency token is twice the size of the input time token due to the representation discussed in Section \ref{section:ew}. The time series encoder only deals with amplitude at given timestamp while the frequency encoder deals with both real and imaginary values at every timestamp, thus making the size of frequency token double. Our arrangement of encoder is inspired by the encoder of Swin-UNet \cite{cao2021swin}. The Swin transformer architecture\cite{liu2021swin}, uses the shifted windows approach, comprising of alternating normal and shifted window transformer blocks. Our encoder differs from the Swin-Unet's encoder in 2 ways. Firstly, Swin-UNet uses a 2D patch merging mechanism, resulting in a 4x downgrade in the number of tokens between consecutive Swin transformer blocks. We use 1D patch merging mechanism, resulting in 2x downgrade in the number of tokens. Secondly, computational complexity of Swin-UNet decreases as depth increases, while we  keep this complexity constant by increasing the dimension of feature maps by a factor of 2 for each patch merging downgrade. The architecture also comprises a bottleneck at the end of both time and frequency encoders for learning of deep feature representation. 

\subsubsection{Decoder}
\label{section:dec}
The decoder has the same dimension  
as that of the encoder, where a patch expanding layer is used in contrast to that of the patch merging layer of the encoders. This patch expanding layer also downgrades the feature map dimension by a factor of 2 in each pass.  Finally, we have a patch expanding layer at the last stage of decoder which aims to reconstruct the feature vector to its original size.
The output of the final patch expanding layer is  considered as the reconstructed waveform. Skip connections are employed between corresponding layers of encoders and the decoder to take advantage of multi-scale feature sets in our architecture.
The role of the decoder is to interpret the learned feature representations from time and frequency domains. There are two ways to combine feature vectors  from the two encoders: One way is to concatenate both the vectors i.e two $W \times H \times D$ 
vectors can be concatenated to get $W \times2H \times D$ or $2W \times H \times D$ as the final representation, while the other way is to stack the feature maps from both the encoders, thus getting the final representation of $W \times H \times 2D$. We  use a stacking mechanism to combine the feature vectors from both time and frequency encoders and hence, the  dimension of decoder in our architecture is double the dimension of either encoder in terms of feature maps.
\begin{figure}[t]
  \centering
  \includegraphics[width=0.47\textwidth]{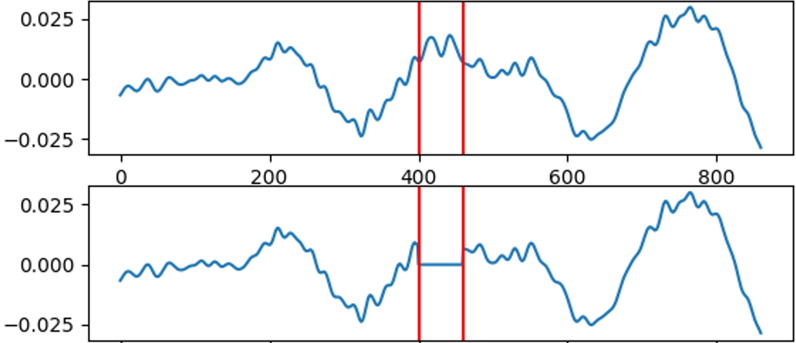}
   \caption{\textbf{Seismic waveform with gap.}The upper graph is the original waveform while the lower graph is the waveform with a random gap}
   \label{fig:seis}
\end{figure}

\begin{figure*}
  \centering
  \begin{subfigure}{0.3\linewidth}
    \centering
    \includegraphics[width = \textwidth]{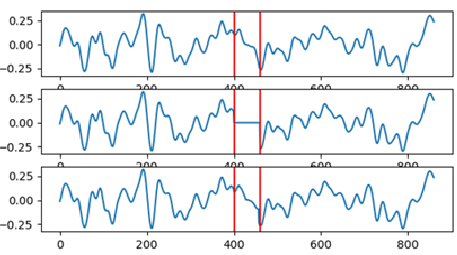}
    \caption{Full Reconstruction}
    \label{fig:favourable}
  \end{subfigure}
  \hfill
  \begin{subfigure}{0.3\linewidth}
    \centering
    \includegraphics[width = \textwidth]{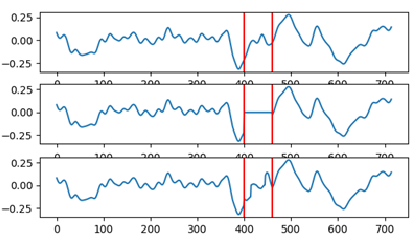}
    \caption{Pattern Detected}
    \label{fig:Pattern}
  \end{subfigure}
  \hfill
  \begin{subfigure}{0.3\linewidth}
   \centering
    \includegraphics[width = \textwidth]{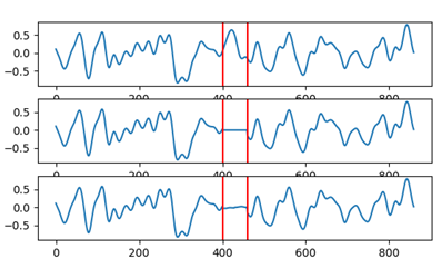}
    \caption{No Reconstruction}
    \label{fig:Failed}
  \end{subfigure}
  \caption{\textbf{Qualitative Results of Waveform reconstruction using Xi-Net.} Three broad categories of results are showcased here. In every figure, the upper graph represents the original waveform, the middle graph shows the random gap, while the bottom graph is the Xi-Net reconstructed waveform. Most of the Xi-Net reconstructed waveforms fall in category (a).}
  \label{fig:short}
\end{figure*}

\section{Experiments}
\label{section:exp}
This section gives an overview of the dataset used in the experiments for evaluation of the performance of Xi-Net with regards to seismic waveform reconstruction and training details concerning the same.

\subsection{Dataset}
\label{sec:data}
The dataset used for these experiments contains 13,696 signals as samples which have been captured by AFTAC\footnote{Data from the TA network were made freely available as part of the EarthScope USArray facility, operated by Incorporated Research Institutions for Seismology (IRIS) and supported by the National Science Foundation, under Cooperative Agreements EAR-1261681} out of which 2,696 are testing samples and 11,000 are training samples . After 
filtering each signal with a Butterworth Bandpass filter \cite{butterworth1930theory} and upsampling which is done by adding 1 point between every two points by taking average. This introduces granularity in the jumps of the waveform in turn helping the model to detect patterns easily. The signal is then pre-processed as discussed in section \ref{section:preprocessing}. We randomly generate gap of 0.5-1 seconds. This data with gaps is used as input into the Xi-Net model while the pre-processed signals without gaps is used as the ground truth (Fig.\ref{fig:seis}). As this is time series data, we only mirror each waveform under augmentation, thus doubling our training set.

\subsection{Training}
We use ImageNet\cite{deng2009large} pre-trained Swin-T \cite{liu2021swin} as the backbone for both encoders and the decoder. Originally Swin shifts the window on image plane (both x and y). We have done our own implementation for such shifting on stream data where shifting will happen in the only axis available. We train our final model using AdamW optimizer for 80 epochs with $10^{-3}$ as the learning rate for the first 50\% of training and dynamically reducing learning rate for the remaining half of the training with weight decay of $10^{-4}$. We use Mean Square Error (MSE) as the loss function. We employ NVIDIA Tesla V100 GPUs for training, and end to end training time was 7 hours on 1 node.

\section{Results and Discussion}
\label{section: rnd}
This section includes the in depth discussion on the context and implications of the results along with the Qualitative and Quantitative analyses of the same. 
\subsection{Qualitative Results}
Here, we visualize three kind of output waveforms. The first type is where the waveform is reconstructed successfully (Fig. \ref{fig:favourable}); the second type is where a pattern is detected but the waveform is not fully reconstructed (Fig. \ref{fig:Pattern}), and lastly, a type where our model fails to reconstruct the signal completely (Fig. \ref{fig:Failed}).  Most of our validation graphs fall in the first category, samples of which can be found \href{https://drive.google.com/drive/folders/1x0m8rGhaWJMaS5u8GsQn9z7pcTBkNb-U?usp=sharing}{here}.

\subsection{Quantitative Results}
In Table \ref{tab:results}, we provide the results of evaluation of our model on four different metrics that are: Discrete Fréchet Distance(DFD)~\cite{eiter1994computing}, Mean Range Difference(MRD)(calculated as the difference between the mean of the ranges of all predicted waveforms, and the mean of the ranges of all original waveforms), Mean Absolute Error(MAE),  and Root Mean Square Error(RMSE). We show the performance of our model in comparison to the original waveform to showcase the quality of reconstruction. For reference, we also include the metric scores of the unfilled gaps. The baseline used for comparison is a 21 layer 1D CNN model implemented by AFTAC, which has been trained for 27 days (5000 epochs).

In general, we see that Xi-Net reconstructs seismic waveforms very satisfactorily, which is visible by the massive difference between the reconstructed metrics and the reference. The improvement seen over the baseline is also noteworthy. Furthermore Xi-Net trains for just 7 hours as compared to the 27 day training time of the baseline, which is a drastic improvement.
\begin{table}[!h]
\centering
\begin{tabular}{lccc}
\toprule
Metric            & Reference(Unfilled) & Baseline & Xi-Net o/p       \\
\midrule
DFD $\downarrow$              & 0.196  & 0.193      & \textbf{0.182}  \\
MRD $\downarrow$            & 0.3467   & 0.2758      & \textbf{0.1861} \\
MAE $\downarrow$              & 0.0955   & 0.0817      & \textbf{0.0813}  \\
RMSE $\downarrow$             & 0.135   & 0.119       & \textbf{0.116} \\
\bottomrule
\end{tabular}
\caption{Performance of Xi-Net model in comparison to the reference(unfilled gaps) and the baseline, across four evaluation metrics. $\downarrow$ : Lower the better}
\label{tab:results}
\end{table}

\section{Ablation Study}
\label{section: abl}
\subsection{Inclusion of Frequency Domain}
The main proposal of our architecture is to combine the benefits of both time and frequency domain. Here, Table \ref{tab:abl1} shows the evaluation of Xi-Net with and without Frequency domain processing. We see that the inclusion of Frequency domain processing improves the performance of the model, as  it facilitates smoother reconstruction and corroborates our initial intuition that both domains are required for good reconstruction. This experiment was performed by running only time series encoder and adjusting the dimensions of the decoder accordingly.

\begin{table}[!h]
\centering
\begin{tabular}{lcc}
\toprule
Metric            & w/o Frequency & w/ Frequency      \\
\midrule
DFD $\downarrow$              & 0.186        & \textbf{0.182}  \\
MRD $\downarrow$            & 0.2238          & \textbf{0.1861} \\
MAE $\downarrow$              &  0.082    & \textbf{0.0813}  \\
RMSE $\downarrow$             & 0.119        & \textbf{0.116} \\
\bottomrule
\end{tabular}
\caption{Comparing variants of Xi-Net with and without frequency domain processing. $\downarrow$ : Lower the better} 
\label{tab:abl1}
\end{table}


\subsection{Single Encoder for Time, Frequency Domain Processing}
The final Xi-Net model with additional training and Time and Frequency domain processing encoders, while still being light, is heavier compared to the model without frequency domain processing, which was implemented earlier. To reduce complexity, an attempt was made to input both time and frequency domain data into a single multi-channel encoder (increasing depth of tokens). Unfortunately the attempt did not fall through and the performance was worse.

\section{Conclusion and Future Work}
\label{section:cfw}
This paper presents Xi-Net, a transformer-based architecture for seismic waveform reconstruction. As it uses a combination of multi-domain data, we believe that Xi-Net can act as a backbone for reconstruction of time series data, in general. Xi-Net also provides the trick to perform self-attention on complex numbers. Although there are not many previous studies in this domain, our proposed Xi-Net has performed satisfactorily on numerous evaluation metrics. While initial results are promising, there is a lot of room for improvement like jitter-free suppression of steep slopes near the gaps’ edges and iterative reconstruction for partially reconstructed samples. An interesting direction to explore is incorporation of a Generative Adversarial Network with our Xi-Net model for refined reconstruction. Additionally, we believe that this combination of self-attention mechanism of transformers, along with the capability to learn using both time and frequency domains can be generalized to several other vision and signal processing tasks.

\section{Acknowledgements}
\label{section:ack}
The authors would like to thank AFTAC (US Air Force) to provide us with the opportunity to work on this project.



\bibliographystyle{IEEEbib}
\bibliography{strings,refs}

\end{document}